\documentclass[aps,twocolumn,showpacs,superscriptaddress,groupedaddress,floatfix, twocolumn,aps,prd,amsmath,amssymb]{revtex4}
\usepackage[english]{babel}
\usepackage[utf8]{inputenc}
\usepackage{epsfig}
\usepackage{multirow}
\usepackage{graphicx}
\usepackage{graphics}
\usepackage{subfigure}
\usepackage{epstopdf}
\usepackage{float}
\usepackage{bm}
\usepackage{braket}
\usepackage{lineno}
\usepackage{slashed}
\usepackage{hyperref}
\usepackage{ulem}
\hypersetup{colorlinks=true, linkcolor=red, filecolor=black, citecolor=blue}

\newcommand{\n}{\noindent}
\usepackage{xcolor}
\usepackage{booktabs}
\usepackage{siunitx}

\begin{document}

\title{Renormalization of the optical band gap through an effective Thirring interaction for massive Dirac-like electrons}

\author{Nilberto Bezerra}
\email{jose.bezerra@icen.ufpa.br}
\affiliation{Faculdade de Física, Universidade Federal do Pará, Avenida Augusto Correa 01, 66075-110, Belém, Pará,  Brazil}

\author{Van Sérgio Alves}
\email{vansergi@ufpa.br}
\affiliation{Faculdade de Física, Universidade Federal do Pará, Avenida Augusto Correa 01, 66075-110, Belém, Pará,  Brazil}

\author{Leandro O. Nascimento}
\email{lon@ufpa.br}
\affiliation{Faculdade de Física, Universidade Federal do Pará, Avenida Augusto Correa 01, 66075-110, Belém, Pará,  Brazil}
\affiliation{Universidade Federal de Campina Grande, Rua Aprígio Veloso 882, 58429-900, Campina Grande, Paraíba, Brazil}

\author{Luis Fernández}
\email{luis.fernandez@ufrontera.cl}
\affiliation{Departamento de Ciencias Físicas, Facultad de Ingeniería y Ciencias, Universidad de La Frontera, Avenida Francisco Salazar 01145, Casilla 54-D, Temuco, Chile}

\date{\today}

\begin{abstract}

We analyze mass renormalization in massive Dirac-like systems in (2+1) dimensions arising from electron-phonon interactions at finite temperatures, employing the large-$N$ expansion. Our model combines the low-energy description of charge carriers in a buckled honeycomb lattice with the low-energy approximation for phonons and electron-phonon interactions in two-dimensional materials. Hence, the system is modeled as a massive Dirac-like field coupled to a two-component vector field $\mathcal{A}_i$, representing the phonon modes. This framework allows us to compute the one-loop electron self-energy at finite temperature, from which we derive the renormalized electronic band gap, $2m^R$. The effective model is subsequently applied to describe the renormalized optical band gap $(E_{\rm{opt}})$ in monolayers of transition metal dichalcogenides (TMDs), including MoS$_2$, MoSe$_2$, WS$_2$, and WSe$_2$, using the relation $E_{\rm{opt}}=2m^R -|E_b|$, where $|E_b|$ is the exciton binding energy that remains constant at the examined temperature. A good agreement is observed with experimental data for reasonable values of the ultraviolet cutoff, $\Lambda \approx 1$~eV. Our main findings indicate that $E_{\rm{opt}}$ remains nearly constant at low temperatures, whereas at higher temperatures it decreases linearly with the temperature $T$. Specifically, we find that $E_{\rm{opt}}$ reduces by approximately $\approx [0.1,0.2]$~eV as the temperature increases from $\approx 4$~K to $500$~K, consistent with recent experimental observations. Furthermore, we estimate the temperature range at which the transition to the linear regime occurs, obtaining typical values within $\approx [110,150]$~K for the four materials under consideration.

\end{abstract}

\maketitle

\section{INTRODUCTION}\label{Sec_Intro}
Four-fermion interactions play a fundamental role in quantum field theory, forming the basis of essential models such as the Thirring model \cite{Thirring}, the Gross-Neveu model \cite{GN}, and the Nambu-Jona-Lasinio (NJL) model \cite{NJL}. These models are of significant importance in both high-energy physics and condensed matter physics. In (3+1) dimensions, the NJL model is widely employed to describe the spontaneous breaking of chiral symmetry, establishing an analogy with the Higgs mechanism in elementary particle theories.

These models provide a theoretical framework for understanding the dynamical mass generation in quarks, a phenomenon analogous to the emergence of a gap in superconductors. They thus offer a suitable platform for exploring non-perturbative effects in quantum field theory. Furthermore, these interactions enable the investigation of various critical phenomena related to parity symmetry breaking and the violation of discrete symmetries.
Additionally, when applied to systems with multiple fermion flavors, these models serve as tools for exploring the mechanism of chiral symmetry breaking in a manner analogous to quantum chromodynamics, yielding a theoretical framework for studying low-energy particle physics phenomena. On the other hand, in the realm of condensed matter physics, conventional superconductivity has been explained in terms of a four-fermion interaction by the BCS (Bardeen–Cooper–Schrieffer) model \cite{MarinoBook}.

In (2+1) dimensions, models with four-fermion interactions acquire new relevance when applied to condensed matter physics, particularly in the study of two-dimensional materials, where the low-energy behavior of charge carriers is often described by a Dirac-like equation. This framework is characterized by two key parameters: the Fermi velocity $v_F$ and the bare mass $m$ \cite{MarinoBook}. The mass parameter $m$ is typically associated with phenomena such as band gaps or spin-orbit coupling \cite{MarinoBook, ZhouBT, Kibis}, both of which play a crucial role in defining the electronic properties of these materials. Moreover, both $v_F$ and $m$ can be renormalized due to electromagnetic interactions, highlighting the importance of considering such effects in theoretical models. 
In particular, the observation of reshaped Dirac cones in graphene \cite{Vozmediano1994, Elias2011}, the experimental observation of the fractional quantum Hall effect in ultraclean samples in graphene \cite{Du, Bolotin, Ghahari, Deam}, and mass renormalization in transition metal dichalcogenides (TMDs) \cite{Luis2020, Bezerra, Excitons2018} have been considered as strong evidence of the relevance of the electronic interactions. In this context, quantum field theory models, such as pseudo-quantum electrodynamics (PQED) \cite{Marino1993} and four-fermion interactions \cite{Luis2020, Charneski}, have been applied to describe the physical properties of two-dimensional materials. Because electrons are charged particles, they naturally interact through the Coulomb potential, which is a physical condition captured by PQED. Nevertheless, due to the many-body interactions of the system, electrons are also subjected to microscopic interactions, such as mechanical vibrations, impurities, and disorder \cite{MarinoBook, Neto2009, PLZhao, BYKHu, JWang, JHChen, Andreas2016, Andreas2020, Luis2020}.

The optical properties of two-dimensional materials have attracted significant attention due to their potential in optoelectronic and photonic applications \cite{Xia, Wang}. One of the main features is the electronic band gap, which plays a crucial role in various technological applications due to the possibility of controlling charge transport properties \cite{Shahbaz, Mandyam, Dutta}. In this case, it is relevant to calculate both the electronic band gap $E_g$ and the optical band gap $E_{\rm{opt}}$, which are closely related. In conventional 3D semiconductors, the optical band gap is typically slightly smaller than the electronic band gap \cite{Luminescence_Book}. The difference between these parameters corresponds to the exciton binding energy $E_b$, which is usually in the order of tens of meV. In general, therefore, we have the relation $E_g=E_{\rm{opt}}+|E_b|$, valid for any temperature (see Fig.~2a in Ref.~\cite{Ugeda}). Furthermore, the electronic band gap can be associated with the mass to the Dirac-like equation through the identity $E_g=2m$. Consequently, following the same approach as before, the optical band gap is determined by $E_{\rm{opt}}= 2m-|E_b|$.

The optical band gap may be understood as the energy required to create an exciton--a bounded electron-hole pair--through the absorption of light in the material \cite{Luminescence_Book, Ugeda}. The electronic band gap is the smallest energy difference between the valence (negative energy) and conduction (positive energy) band of the charge carrier. The binding energy, on the other hand, is due to the Coulomb attraction between electrons and holes. It is known, however, that both the electronic and optical band gaps are modified, not only by the intrinsic structure of the material but also by external factors such as temperature \cite{Jung, Hanbicki}. Furthermore, at least two main processes exist for generating a temperature dependence in these band gaps of the material, namely the lattice expansion and electron-phonon (el-ph) interactions \cite{Vina, Yiming}. 
Phonons are quantized quasiparticles representing mechanical vibrations in material and play a crucial role in determining various properties in solids, such as optical and thermal responses \cite{Amigo}. In two-dimensional crystals, like graphene and transition metal dichalcogenides, unique phonon modes--particularly high-frequency, out-of-plane phonons--strongly influence thermal conductivity and charge carrier mobility.
These phonons interact with electronic states, impacting the overall thermal management and electronic efficiency of these materials. At finite temperatures, thermal fluctuations can induce structural deformations in the lattice, modifying phonon transport and, thus, affecting the thermal and electronic properties of these 2D materials \cite{Voz2010, Neto2009, Nika}. Notably, the exciton binding energy in two-dimensional semiconductors tends to be higher than in its three-dimensional counterparts, often reaching values in the order of hundreds of meV \cite{Hanbicki, Ugeda}. This characteristic highlights the distinct quantum confinement effects present in 2D materials, which can significantly alter their electronic and optical behavior in response to temperature \cite{Komsa}.

In three-dimensional semiconductors, the temperature dependence of the band gap has been largely discussed in literature since the 1950s \cite{Fan, Bardeen}. Furthermore, the optical and electronic band gaps, depending on the temperature, are calculated with the same models, because of their small energy difference (the binding energy of the exciton). Recently, for two-dimensional semiconductors, the same renormalization for both band gaps has been observed in experimental works \cite{Jung, Stevens, Liu, Choi, Tongay}. The main conclusion is that the two types of band gap decrease as we increase the temperature from approximately 4~K to 500~K. 
Although the experimental findings have been explained by phenomenological equations \cite{Varshni, Bludau, Vina, ODonnell}, it is expected that such an effect would be driven by the el-ph interaction \cite{Bludau, Vina, ODonnell}. Here, we shall propose a physically motivated model for describing this band gap renormalization in some TMDs using an approximated version of the full el-ph interactions, within an effective quantum field theory model. 
It is interesting to note that the temperature dependence of both the optical and electronic band gap is the same \cite{Jung, Hanbicki}, despite the notably high exciton binding energy in two-dimensional semiconductors. Indeed, for temperatures up to 350~K, the experimental data shows that the exciton binding energy remains almost unchanged by the temperature of the thermal bath \cite{Jung}. 
The reason behind such behavior is likely to be that the exciton binding energy and the electronic band gap are strongly sensitive to the strength of the Coulomb interactions rather than temperature \cite{Raja, Ataei, Ugeda}. 
Nevertheless,  the optical band gap is less sensitive to the Coulomb potential and more sensitive to temperature effects \cite{Ugeda, Qiu}. This, as it will be clear later, is an important assumption for our main result.  

The experimental measurements of the renormalized band gaps are obtained due to the strong light-matter interaction, hence these can be measured through spectroscopic techniques \cite{Paul, Hanbicki, Ugeda, Steinhoff, Liu}. Therefore, it is crucial to develop a better understanding of carrier scattering, including the intrinsic contributions from el-ph interactions. These interactions arise from changes in the atomic potential caused by atomic vibrations within the lattice \cite{Sasaki_1, Sasaki_2}. In TMDs monolayers, the optical phonon modes are expected to induce an el-ph scattering \cite{Paul, Lee, Sohier}, where the optical phonon branch is particularly relevant to the direct band gaps at the $K$ and $K'$ points in the first Brillouin zone \cite{Jung, Hanbicki, Ugeda}. Unfortunately, a full model for describing the el-ph interaction is not known yet for any energy scale. However, low-energy approximations can be conveniently made within an effective model using quantum field theory methods \cite{Voz2010}. 

In this work, we consider a low-energy effective model for describing el-ph interaction in two-dimensional materials with a buckled honeycomb lattice. This lattice implies that the quasiparticles are described by a massive Dirac-like equation. The effective description of the phonon field yields a two-component vector $\mathcal{A}_i$, where $i=1,2$ is related to the two possible polarizations. The interaction term, within this regime, is given by a three-point vertex interaction and gives a one-loop quantum correction to the band gap of the matter field. For calculating this correction we use the large-$N$ expansion and include the thermal bath, using the imaginary-time formalism. Thereafter, we obtain the renormalized band gap $2m^R(T,\Lambda)$ as a function of temperature, through the Schwinger-Dyson equation for the corrected propagator of the matter field. Because the model is only defined for a finite ultraviolet cutoff $\Lambda$, hence it does not have the typical problems of divergencies, such as in quantum electrodynamics. 
This parameter is, essentially, the electronic energy at the boundary of the Brillouin zone and our field-theory approach describes the low-energy dynamics of the electronic quasiparticles in the TMDs.
Indeed, our approach should be understood as an effective model of the full theory for the complete el-ph interaction, which allows us to avoid the hurdle of working on a complicated many-body problem. Thereafter, we compare our expression for $E_{\rm{opt}}(T,\Lambda)=2m^R(T,\Lambda)-|E_b|$ with the experimental findings for monolayers of TMDs obtained in Ref. \cite{Liu}. After fixing three physical parameters, namely, the bare mass ($m$), the energy cutoff ($\Lambda$), and the coupling constant ($g$),  we show that our result is in good agreement with the experimental findings for four different types of two-dimensional materials. 

This work is divided as follows: In Sec.~\ref{Sec_FeynmanRules}, we present our model and its Feynman rules. In Sec.~\ref{Sec_Self-energy}, we calculate the electron self-energy at a finite temperature. In Sec.~\ref{Sec_BGR}, we calculate the renormalized mass and compare it with experimental results. 
In Sec.~\ref{Sec_SummOut}, we summarize and discuss our main results. 
We also provide in Appendix~\ref{App_Model} the construction of the model, while Appendices~\ref{App_Limits} and \ref{App_SEwithoutTemp} provide additional details of the calculations for the zero and high temperature, respectively.

\section{The effective model and the Feynman rules}\label{Sec_FeynmanRules}

We consider a low-energy effective model for describing the el-ph interaction in two-dimensional materials with a band gap in its energy spectrum.
In two-dimensional materials with a buckled honeycomb lattice, the charge carriers, i.e., the quasiparticles, obey a Dirac-like equation, where the mass describes the energy band gap of the electron \cite{MarinoBook, Neto2009}. 
The phonon field in these materials can be represented as a vector field associated with the ion displacements, capturing the lattice vibrations and their coupling to the electronic states \cite{Sasaki_1, Sasaki_2, Ando_1, Ando_2} 
(for more details on the model approximations, see Appendix~\ref{App_Model}). Therefore, 
based on these considerations, the Lagrangian density for the system reads
\begin{equation}\label{L_Model}
\begin{split}
	\mathcal{L} =\bar{\psi}_\mathtt{a} \left(i \gamma^{\mu} \partial_{\mu} - m -\frac{\gamma^i \mathcal{A}_i}{\sqrt{N}} \right)\psi_\mathtt{a} + \frac{\mathcal{A}^i \mathcal{A}_i}{2 g}  \, ,
\end{split}
\end{equation}
where the spinor field reads $\psi_{\mathtt{a}}(\bold r) = \left(\hat{A}_{\mathtt{a}} \, \hat{B}_{\mathtt{a}} \right)$ and $\psi_{\mathtt{a}}^{\dagger}(\bold r) = \left(\hat{A}_{\mathtt{a}}^* \, \hat{B}_{\mathtt{a}}^* \right)^{T}$. $\hat{A}_{\mathtt{a}}$ ($\hat{B}_{\mathtt{a}}$) and $ \hat{A}_{\mathtt{a}}^{\dagger}$ ($ \hat{B}_{\mathtt{a}}^{\dagger}$) are the annihilation and creation operators of electrons (holes) in the two inequivalent sublattices $A$ and $B$ of the honeycomb lattice in real space, respectively. 
$\gamma^\mu$ are the rank-2 Dirac matrices that obey the anticommutation relation, given by $\{\gamma^ \mu, \gamma^\nu \} = -2 \delta^{\mu \nu}$, where $\delta^{\mu \nu}\equiv {\rm diag}(1,1,1)$ is the Euclidean metric. 
The Greek index read $\{0,1,2\}$ and the differential operator is defined as $ \partial_\mu = (\partial_0, v_F\partial_i)$, where $v_F$ is the Fermi velocity of the quasiparticles. 
The flavor index $\mathtt{a}=\{ K \uparrow, K\downarrow, K'\downarrow, K' \uparrow \}$ specifies the valley and the spin to which the electron (hole) belongs, respectively, thereby $N=4$ is the total degree of freedom \cite{PRX}. 
$\mathcal{A}_i$ is the phonon field, where the Latin indexes are given by $\{ 1, 2 \}$. It is important to emphasize that our model neglects phonon dynamics, as their influence is significantly less relevant compared to the dynamics of charge carriers (see Appendix~\ref{App_Model}). 
The Thirring interaction is straightforwardly obtained as we integrate out $\mathcal{A}_i$ in Eq.~(\ref{L_Model}) Ref.~\cite{Andreas2016, MGomes}.
$g$ is our coupling constant and has units of the inverse of mass \cite{Andreas2020}. Within our low-energy model, as mentioned before, $m$ describes the bare band gap of the charge carriers.
Indeed, we shall consider an ultraviolet cutoff in the model, given by $\Lambda=\hbar v_F/a$ \cite{Vozmediano1994, Excitons2018}, where $a$ is the lattice parameter. We shall use $c=\hbar=k_B=1$ everywhere, but recover physical units when it is convenient.

The bare fermion propagator for the matter field, in Eq.~\eqref{L_Model}, is given by
\begin{equation}\label{Ferm_Propagator}
S_{0F} (p) = \frac{-1}{p^\mu \gamma_\mu - m} \, ,
\end{equation}
where $p_\mu=(p_0,v_F \bf p)$ and its pole yields the dispersion relation of the quasiparticle, given by $p_0 \equiv E(\textbf{p})= \pm \sqrt{v_F^2 \textbf{p}^2 + m^2}$. This is the main connection between the Dirac-like equation with the low-energy model for quasiparticles in two-dimensional materials. The bare phonon propagator reads
\begin{equation}\label{Phonon_Prop}
\Delta^{(0)}_{i j} (p) = g \, \delta_{i j},
\end{equation}
which has no dynamics at tree-level approximation. In this case, the dynamics of phonons in the lattice are neglected as they are less relevant than electronic dynamics. Finally, 
\begin{equation}
\Gamma^i = -\frac{\gamma^i}{\sqrt{N}}
\end{equation}
is the vertex interaction.

In the next section, we will discuss the effect of introducing the vacuum polarization tensor in the calculation of the electron self-energy and calculate this at finite temperatures.

\section{The Electron self-energy at Finite Temperature In Large-N approximation }\label{Sec_Self-energy}

Here, we consider the large-$N$ approximation while $g$ is fixed \cite{Coleman}. In (2+1) dimensions, the Thirring model is perturbatively non-renormalizable \cite{MGomes}. However, the large-$N$ approximation modifies the Thirring interaction as a trilinear interaction with a perturbative parameter given by 1/$N$ to the electron self-energy. Another reason why we choose the large-$N$ expansion is for the sake of comparison with the electron-electron interaction in two-dimensional materials. 
In this case, the fine-structure constant may be large, unless one considers a strong screening of the Coulomb potential, which is usually provided by putting the material above a substrate with a high dielectric constant \cite{YCho}. 
This somehow poses a condition for the perturbative approximation. The large-$N$ expansion is a formal method that allows us to circumvent this problem. However, qualitative results, such as the Fermi velocity renormalization, seem to indicate that both methods provide similar conclusions. In our calculation, we expect that our results are also confirmed either for a small coupling between the electron and the phonon or when $N$ is large. This, however, is more subtle because the physical value of $N$ is four [related to the valley and the spin that belongs to the electron (hole)]. Hence, it seems safer to assume that the el-ph coupling is small, such that the one-loop approximation yields a reasonable result \cite{Kumar}. In this regime our results are correct, otherwise one should include higher-order corrections. 

\begin{figure}[h!]
	\centering
	\includegraphics[width=0.45\textwidth]{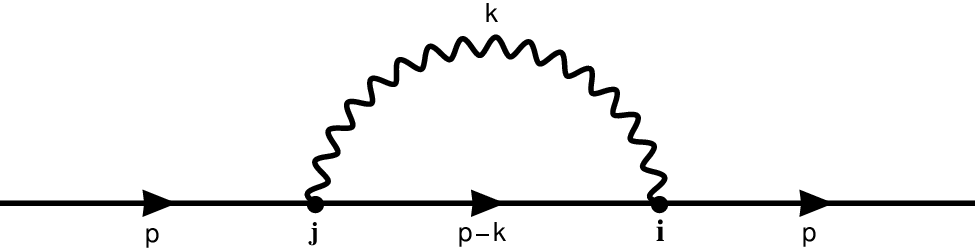}
	\caption{ The electron self-energy. The straight line represents the fermion propagator while the waved line denotes the phonon propagator.}\label{Fig_SE}
\end{figure}

In this case, in order to compute the electron self-energy at the lowest order in the coupling constant $g$, we must calculate the Feynman diagram represented by Fig.~\ref{Fig_SE}. The electron self-energy is given by
\begin{equation}\label{SE_T0}
\Sigma (p) = \int \frac{d^3 k}{(2 \pi)^3} \Gamma^i S_{0 F}(p-k) \Gamma^j \Delta_{ij} (k) \,,
\end{equation}
\n where $\Delta_{ij} (k)$ is the full phonon propagator. Furthermore, for comparison with experimental measurements, it is more interesting to consider a system in equilibrium with a thermal bath. Hence, we apply the imaginary-time formalism to introduce the effects of finite temperature \cite{Matsubara, Das_Book}. In this formalism, we transform the integral in $k_0$ into a sum over the Matsubara frequencies, i.e.,
\begin{equation}
\int dk_0 \, f(k_0,\textbf{k}) \to \frac{2 \pi}{\beta}\sum_{n= - \infty}^{\infty} f(\omega_n, \textbf{k}) \, ,
\end{equation}
where $f(k_0,\textbf{k})$ is an arbitrary integrand and $\beta^{-1} = T$ is the equilibrium temperature. Moreover, the three-momentum is  $(k_0, \textbf{k}) \to (\omega_n, \textbf{k} )$ and $(p_0, \textbf{p}) \to (\omega_l, \textbf{p})$, with $\omega_x = (2x + 1)\pi /\beta $ for fermions and $\omega_x = 2x \pi /\beta $ for bosons. $\omega_x$ is the Matsubara frequency, and the label $x= \{n, l\}$ is its mode of vibration.

Before we calculate the electron self-energy, we first discuss the effect of incorporating the vacuum polarization tensor into the computation of the full-phonon propagator at finite temperatures. In this case, the full-phonon propagator is given by the Schwinger-Dyson equation, namely,
\begin{equation}
\Delta^{-1}_{ij}(p_0,\textbf{p},T)=(\Delta^{(0)}_{ij})^{-1}-\Pi_{ij}(p_0,\textbf{p},T), \label{SDE}
\end{equation}
\n where $\Pi_{ij}(p_0,\textbf{p}, T)$  are the standard spatial components of the vacuum polarization tensor of quantum electrodynamics in (2+1)D at finite temperature \cite{Dorey}. Within the static limit $p_0=0$, and for $\textbf{p} \rightarrow 0$, we can obtain
\begin{equation}
	\Pi_{ij}(T,\textbf{p}) = -\Pi(T,\textbf{p})\, \,P_{ij},
\end{equation}
where
\begin{equation}
\Pi(T,\textbf{p})=\frac{\textbf{p}^2}{12\pi} \frac{\tanh\left( \frac{m}{2T} \right)}{m}\, 
\end{equation}
and $P_{ij}=\delta_{ij}-\frac{p_{i}p_{j}}{\textbf{p}^2}$ is the transversal projection operator. Similar results for massless fermions can be found in Refs.~\cite{Dorey, Perez}. We can invert Eq.~(\ref{SDE}) and, using $\Delta^{-1}_{ij}\,\Delta^{jk}=\delta^k_i$, we obtain
\begin{equation}\label{ApB3}
	\Delta_{ij}(T,\textbf{p})= \left[\frac{g}{1+ g \, \Pi(T,\textbf{p}) }\right]\left[\delta_{ij}+ g \, \Pi(T,\textbf{p}) \frac{p_i p_j}{\textbf{p}^2}\right].
\end{equation}

Therefore, using Eq.~(\ref{ApB3}), it follows that $\Delta_{ij}\rightarrow g \delta_{ij}$ when $g\,\Pi(k_BT,\textbf{p}) \ll 1$, where we have recovered the physical units of temperature by performing $T\rightarrow k_B T$. The central idea of our approximation is to neglect any correction originating from the phonon self-energy. Indeed, this is necessary for neglecting the dynamical effects of the phonon field, such that our initial model in Eq.~(\ref{L_Model}) is still reliable. Hence, according to Eq.~(\ref{ApB3}), our results are in agreement with our initial assumptions and further corrections are expected to be quite small as long as one neglects the dynamics of phonons. Furthermore, the static limit $p_0=0$ makes sense when the electron velocity $v_F$ is much less than the light velocity $c$. 
It is worth mentioning that the vacuum polarization effect on the auxiliary field propagator, $\Delta_{\mu\nu}(p)$, was investigated in Ref.~\cite{MGomes} in the context of dynamical chiral symmetry breaking for a Lorentz-invariant version of the Dirac Lagrangian.


Using these considerations, we can write Eq.~\eqref{SE_T0} as 
\begin{equation}
\begin{split}\label{SE_T}
\Sigma_l (T,\bold p) =& \frac{2 g T}{N} \int \frac{d^2 k}{(2 \pi)^2} \sum_{n=-\infty}^\infty \gamma^i  \\
& \times \frac{ \gamma^0 [1-2(n-l)] \pi T + v_F \pmb{\gamma}.(\textbf{p} - \textbf{k}) + m }{ [1-2(n-l)]^2 \pi^2 T^2 + E^2(\textbf{p}, \textbf{k},m)}\gamma_i\,,
\end{split}
\end{equation}
where $E(\textbf{p}, \textbf{k},m) = \sqrt{v_F^2 |\bold p - \bold k|^2 + m^2 }$ and $\pmb{\gamma}.\textbf{p} = \gamma^i p_i$. 
Eq.~\eqref{SE_T0} is not exactly the same as Eq.~\eqref{SE_T}, but it is an $l$-component. 
However, the most dominant vibration mode is $l=0$ \cite{Leandro}. This vibrational-mode approximation remains consistent within the framework of the local approximation \cite{Chubukov}, where the velocity of the boson is much less than the velocity of the fermion, which is in agreement with our
approximations. Hence, we can consider $p_0=(2l+1)\pi/\beta\to~\pi/\beta$ \cite{Chubukov} and, for the sake of simplicity, we use $\Sigma(p)\to\Sigma_0(T,\bold p)=\Sigma(T,\bold p)$. Using this, we have
\begin{equation}
\Sigma (T,\bold p) = \frac{2 g T}{N} \int \frac{d^2 k}{(2 \pi)^2} (\gamma^0 \pi T \, S_1 - m\, S_2 )\,,
\end{equation}
where the sums $S_1$ and $S_2$ are given by
\begin{equation}\label{Sum_S1}
S_1 = \sum_{n=-\infty}^\infty \frac{(1-2n)}{(1-2n)^2 \pi^2 T^2 + E^2(\bold p,\bold k, m )} = 0\,,
\end{equation}
and
\begin{equation}
\begin{split}\label{Sum_S2}
S_2 & = \sum_{n=-\infty}^\infty \frac{1}{(1-2n)^2 \pi^2 T^2 + E^2(\bold p,\bold k, m )} \\
&= \frac{1 - 2 n_F ( \bold p,\bold k, m, T)}{2 \, T \, E(\bold p,\bold k, m ) }\,,
\end{split}
\end{equation}
with
\begin{equation}
n_F (\bold p, \bold k, m, T) = \frac{1}{ \exp \left( \frac{E(\bold p, \bold k, m )}{T}\right) + 1}\,,
\end{equation} 
which is the Fermi-Dirac distribution. The solutions of $S_1$ and $S_2$ in Eqs.~\eqref{Sum_S1} and \eqref{Sum_S2} can be found in Ref.~\cite{Bezerra}. Note that the term proportional to $\pmb \gamma$ in Eq.~(\ref{SE_T}) vanishes due to the product $\gamma^i \gamma^j \gamma_i = 0$.

Next, we solve the loop integral by using polar coordinates, where $d^2 k = \mathrm{k}\, d\theta \, d\rm k$, $|\bold k | = \rm k$, and $|\bold p | = \rm p$. Furthermore, we perform the following substitutions: $\mathrm k  \to \mathrm{(k+p)}/v_F$ and $\mathrm p  \to\mathrm p/v_F$. Hence, after some algebra, we find
\begin{equation} \label{SE_Tint}
\Sigma (T,\Lambda) = - \frac{g \, m }{N} \left( I_1 - 2 I_2 \right)\, ,
\end{equation}
where
\begin{equation}
\begin{split}\label{Inte1}
I_1 =& \frac{1}{2 \pi v_F^2} \int^\Lambda_0 d\, \mathrm k \frac{\rm k}{(\mathrm k^2 + m^2)^{1/2}}\\
=& \frac{1}{2 \pi v_F^2} \left[ (\Lambda^2 + m^2)^{1/2}  - m \right]\,,
\end{split}
\end{equation}
and
\begin{equation}
\begin{split} \label{Inte2}
I_2 =& \frac{1}{2 \pi v_F^2} \int^\Lambda_0 d\, \mathrm k \frac{\rm k}{(\mathrm k^2 + m^2)^{1/2}} n_F (\mathrm{k}, m, T ) \\
=& - \frac{T}{2 \pi v_F^2} \ln \left[ \frac{1+\exp\left( -\frac{ (\Lambda^2 + m^2)^{1/2}}{T} \right)}{ 1 + \exp\left(- \frac{m}{T} \right)} \right] \,.
\end{split}
\end{equation}
The ultraviolet cutoff $\Lambda$ is an energy scale that is inversely proportional to the lattice parameter $a$, i.e. $\Lambda = \hbar \, v_F/a$ \cite{Vozmediano1994, Excitons2018}. 

Finally, using Eqs.~(\ref{Inte1}) and (\ref{Inte2}) into Eq.~(\ref{SE_Tint}), the electron self-energy reads
\begin{equation}
\begin{split}
\Sigma (T, \Lambda) =& - \frac{g\, m}{2\pi  N  v_F^2} \left\lbrace (\Lambda^2 + m^2)^{1/2}  - m \right.\\
&+ \left. 2T \ln \left[ \frac{1+\exp\left( - \frac{(\Lambda^2 + m^2)^{1/2}}{T} \right)}{ 1 + \exp\left(- \frac{m}{T} \right)} \right] \right\rbrace \, . \label{SE_TFinal1}
\end{split}
\end{equation}
The low- and high-temperature limits are easily calculated when we use the logarithm property $\ln(xy) = \ln(x)+\ln(y)$ and the identity $2 \cosh(z) = \exp(z)+\exp(-z)$ in Eq.~(\ref{SE_TFinal1}). Hence, after a straightforward algebra, we have
\begin{equation}\label{SE_TFinal2}
\Sigma (T, \Lambda) = - \bar{g}\, m \, T \ln \left[ \frac{\cosh\left( \frac{(\Lambda^2 + m^2)^{1/2}}{2T} \right)}{ \cosh\left( \frac{m}{2T} \right)} \right],
\end{equation}
where $\bar g = g/(\pi N v_F^2)$ is our new free parameter. Note that Eq.~\eqref{SE_TFinal2} is the exact electron self-energy for the model in Eq.~\eqref{L_Model} (which has been obtained after some approximations) and it holds for any temperature. Therefore, when the temperature tends to zero, the temperature dependent-term vanished in Eq.~(\ref{SE_TFinal1}) (see Appendix~\ref{Sub_App_T0} for more details), i.e., we obtain only the vacuum correction to the electron propagator [see Eq.~\eqref{SE_Solv_T0}]. 
On the other hand, when considering the ultra-high-temperature limit $T \gg \Lambda$ (in practice, $T \to \infty$), the electron self-energy vanishes (see Appendix~\ref{Sub_App_Tinfinity}), and the electron mass is not normalized. 
In the context of TMDs, this temperature should be close to the so-called decomposition temperature ($T_{\rm{dec}}$), where the lattice structure of the material becomes unstable and decomposes, for example, $T_{\rm{dec}}\approx 993.0$~K for MoSe$_2$ monolayer \cite{Choi}. Obviously, our model is not accurate in this regime.

\section{The Band Gap Renormalization}\label{Sec_BGR}

In this section, we calculate the mass renormalization due to the electron self-energy at finite temperature and associate it with the optical band gap $E_{\rm opt}$. Thereafter, we compare $E_{\rm opt}$ in the following TMDs: MoS$_2$, MoSe$_2$, WS$_2$, and WSe$_2$.

\subsection{The Renormalized Mass}

We start with the Schwinger-Dyson equation for the full electron propagator $S_F(T, \mathrm p)$. This is given by
\begin{equation}
S^{-1}_{F} (T, \mathrm p) = S^{-1}_{0F} (p) - \Sigma (T, \Lambda),  \label{Rfermprop}
\end{equation}
where $S^{-1}_{F} (T, \mathrm p) = \gamma^0 \omega + v^R_F \gamma^i p_i - m^R $ is the corrected propagator within our approximation. $S^{-1}_{ 0F}$ and $\Sigma (T, \Lambda)$ are given by Eqs.~\eqref{Ferm_Propagator} Eq.~\eqref{SE_TFinal2}, respectively.
For the sake of comparison with the experimental findings, we recover the explicit Boltzmann constant: use $T \to k_B T$ ($k_B=8.617 \times 10^{-5}$ eV.K$^{-1}$).  Having this in mind,  the renormalized mass is given by
\begin{equation}\label{mr1}
m^R (T,\Lambda) = m \left[ 1 + \bar{g} \, k_B T \ln \left[ \frac{\cosh\left( \frac{(\Lambda^2 + m^2)^{1/2}}{2 k_B T} \right)}{ \cosh\left( \frac{m}{2 k_B T} \right)} \right] \right],
\end{equation}
where $m^R$ is the renormalized mass and $m$ is the bare mass, both measured in eV as long as the energy scale $\Lambda$ is fixed in eV. Here, it is worth mentioning that $m_R$ is half of the fundamental band gap $E_g$. Furthermore, it can be related to the effective mass obtained from the band curvature. We shall discuss this in detail at the end of Sec.~\ref{SubSec_Camparasion}. In the next section, we shall extract the experimental data for the optical band gaps of the aforementioned TMDs from Ref.\citep{Liu}. 
These data are analyzed with the help of Eq.~\eqref{mr1}.

Eq.~\eqref{mr1} also may be written as
\begin{equation}\label{mr2}
\begin{split}
m^R(T,\Lambda) &= M(0,\Lambda) + \bar{g} \, m \, k_B T \\
&\times \ln \left[ \frac{\exp \left( \frac{2}{\bar{g} \, k_B T}\frac{m-M(0,\Lambda)}{m} \right) + \exp\left( \frac{m}{k_B T} \right)}{\exp\left( \frac{m}{k_B T} \right) + 1}\, \right],
\end{split}
\end{equation}
where
\begin{equation}\label{M0}
M(0,\Lambda) = m + \frac{\bar{g}}{2} m \left[ (\Lambda^2 + m^2)^{1/2} - m \right].
\end{equation}
In both cases, for consistency, we show that both Eq.~\eqref{mr1} and Eq.~\eqref{mr2} yield $m^R(T\to 0,\Lambda) = M(0,\Lambda)$ when we consider $T \to 0$ [see Eq.~\eqref{mr0}].

It is worth mentioning that, unlike \textit{ab initio} methods, our result is an analysis that examines the influence of the el-ph interaction on the renormalized mass as a function of temperature. Consequently, at $T=0$, we can not predict either the zero-point renormalization \cite{Feliciano, Ortenzi, Miglio, Roy} for the mass, which reduces the bare energy gap, or other parameters depending on details of the band structure of the material.

\subsection{Comparison with experimental data}\label{SubSec_Camparasion}

In this section, we use our theoretical result to compare with the temperature-dependent optical band gaps measured in Ref.~\cite{Liu}. Note that our result has three free parameters, namely, $\Lambda$, $m$, and $\bar{g}$ that should be fixed. It is important to emphasize that the empirical equations, such as the Varshni equation \cite{Varshni} and other proposals based on the Bose-Einstein statistical factor for the phonon emission and absorption \cite{Vina, ODonnell} also have three free parameters. 

Firstly, we relate the renormalized electronic band gap to the optical band gap. This can be made through the exact equation, given by $E_{\rm{opt}}(T,\Lambda)=E_g(T,\Lambda)-|E_b(T,\Lambda)|$ \cite{Ugeda, Hanbicki, Luminescence_Book}, as it has been discussed in Sec.~\ref{Sec_Intro}. As we have derived below from Eq.~(\ref{Ferm_Propagator}), the two-band energy of our free charge carriers reads $E_\pm(\textbf{p})=\pm\sqrt{v_F^2\textbf{p}^2+m^2}$. Hence, the electronic band gap is $E_g=|E_+(\textbf{p}=0)-E_-(\textbf{p}=0)|=2m$, for $m\geq 0$ at zero temperature. Our renormalized propagator in Eq.~(\ref{Rfermprop}) yields a similar result, with $E^R_g(T,\Lambda)=|E^R_+(\textbf{p}=0)-E^R_-(\textbf{p}=0)|=2 m^R(T,\Lambda)$, where $m^R(T,\Lambda)$ is given by Eq.~\eqref{mr1}. Note that this is only the simplest renormalization condition, where the pole of the propagator is the physical mass of the particle described by the corresponding field. 

\begin{table}[h]
            \centering
            \caption{Fitted parameters for TMD monolayers.}
            \begin{tabular}{lccc}
            \hline
                \bf{Material} & $\Lambda$(meV) & $m$(meV) & $\bar{g}$(eV$^{-1}$) \\
                \hline
                \bf{MoS$_2$} & 956.9 & 12.24 & 221.5 \\
                \bf{MoSe$_2$} & 849.9 & 9.50 & 263.5 \\
                \bf{WS$_2$} & 828.3 & 11.74 & 259.9 \\
                \bf{WSe$_2$} & 727.9 & 10.34 & 281.9 \\
                \hline
            \end{tabular}\label{Table_1}
\end{table}

Next, motivated by the experimental results in Ref.~\cite{Jung}, we assume that the binding energy remains nearly constant as the temperature of the thermal bath increases. In other words, we consider the exciton binding energy to be temperature-independent, i.e., $|E_b(T=0,\Lambda)|=E_b(\Lambda) > 0$. Furthermore, the exciton binding energies for MoS$_2$, MoSe$_2$, WS$_2$, and WSe$_2$ have been determined both experimentally \cite{Ugeda, Hill, He, Chernikov} and theoretically \cite{Excitons2018, Timothy, Ilkka}, with their values potentially varying due to substrate screening effects \cite{Ugeda, Hill, He, Chernikov}. Among the reported measurements, we fix the exciton binding energies of MoS$_2$ and MoSe$_2$ at $E_b^{\rm MoS_2}=E_b^{\rm MoSe_2}=500.0$~meV, and for WS$_2$ and WSe$_2$ at $E_b^{\rm WS_2}=E_b^{\rm WSe_2}= 400.0$~meV, values which are consistent with experimental observations. In order to fit the experimental data, we identify the optical band gap $E_{\rm{opt}}$ as our renormalized $E^R_g$, hence,
\begin{equation}\label{Eopt}
\begin{split}
E_{\rm opt} (T,\Lambda) =&  -E_b + 2m  + 2m \bar{g} \, k_B T \\ 
& \times\ln \left[ \frac{\cosh\left( \frac{(\Lambda^2 + m^2)^{1/2}}{2 k_B T} \right)}{ \cosh\left( \frac{m}{2 k_B T} \right)} \right] .
\end{split}
\end{equation}
	
\begin{figure}[h!]
	\centering
	\includegraphics[width=0.48\textwidth]{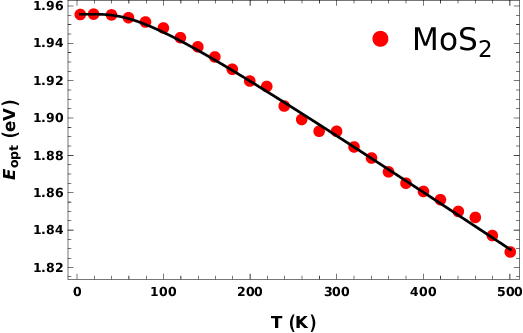}
	\caption{Red circles are experimental data taken from Ref.~\cite{Liu}. The black curve was plotted from Eq.~\eqref{Eopt}. For the best fit we have $\Lambda= 956.9$~meV, $m= 12.24 $~meV, and $\bar g= 221.5 $~eV$^{-1} $.}\label{Fig_MoS2}
\end{figure}

\begin{figure}[h!]
	\centering
	\includegraphics[width=0.48\textwidth]{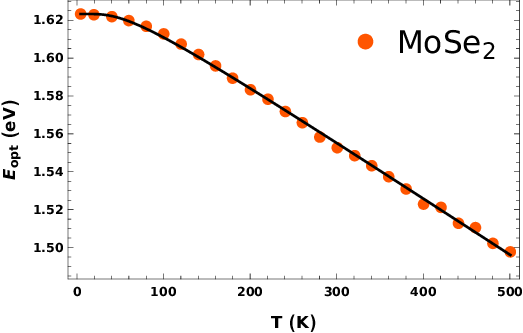}
	\caption{Orange circles are experimental data taken from Ref.~\cite{Liu}. The black curve was plotted from Eq.~\eqref{Eopt}. For the best fit we have $\Lambda= 849.9$~meV, $m= 9.50$~meV, and $\bar g= 263.5 $~eV$^{-1} $. }\label{Fig_MoSe2}
\end{figure}

\begin{figure}[h!]
	\centering
	\includegraphics[width=0.48\textwidth]{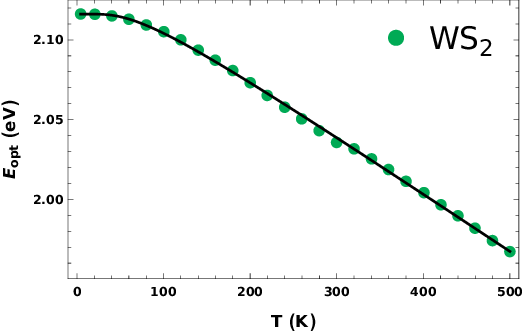}
	\caption{Green circles are experimental data taken from Ref.~\cite{Liu}. The black curve was plotted from Eq.~\eqref{Eopt}. For the best fit we have $\Lambda= 828.3$~meV, $m= 11.74 $~meV, and $\bar g= 259.9 $~eV$^{-1} $.}\label{Fig_WS2}
\end{figure}

\begin{figure}[h!]
	\centering
	\includegraphics[width=0.48\textwidth]{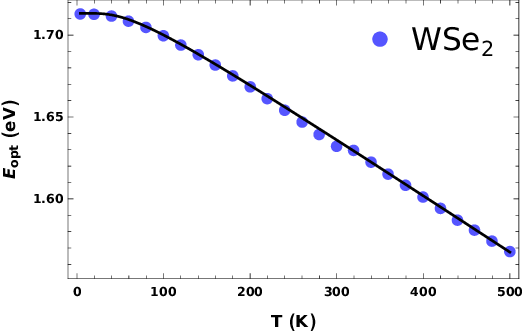}
	\caption{Blue circles are experimental data taken from Ref.~\cite{Liu}. The black curve was plotted from Eq.~\eqref{Eopt}. For the best fit we have $\Lambda= 727.9$~meV, $m= 10.34 $~meV, and $\bar g= 281.9 $~eV$^{-1} $.}\label{Fig_WSe2}
\end{figure}

The comparisons made, from Figs.~\ref{Fig_MoS2} to \ref{Fig_WSe2}, show the renormalized optical band gap of some two-dimensional materials, as a function of temperature. 
The dots are the experimental data and the continuous line is our result, obtained in Eq.~\eqref{Eopt}. The cutoff is proportional to the Fermi velocity $v_F$ and inversely proportional to the lattice parameter $a$. After comparing our theoretical result with the experimental points, the values obtained for $\Lambda$ are close to the value obtained in Ref.~\cite{Excitons2018}, i.e., $\Lambda \approx 1$~eV. Furthermore, it is well known from the literature that the lattice parameter of these TMDs is around 3~\AA ~\cite{Peng, Mathew}, and the Fermi velocity is in order of $10^5$~m/s \cite{Xiao, Hao, Mitioglu}. Using these numbers and $\Lambda=\hbar v_F/a$, we conclude that our estimated cutoff has a reasonable value and does not change drastically for different materials. This number also has an important feature as it represents an upper limit for the usefulness of the Dirac approximation, because when $p\approx \Lambda$ and if $\Lambda$ is much larger than $1$~eV, then some high-energy corrections should be considered for describing the charge carriers. After considering these limitations, we conclude that our result is in good agreement with the experimental data and, therefore, the proposed model seems to effectively capture all the main physics, concerning the el-ph interactions, that describe the renormalization of the optical band gap in these TMDs. In Table~\ref{Table_1}, we show the parameters used for the materials when we compare Eq.~\eqref{Eopt} with the experimental data.
	
From Figs.~\ref{Fig_MoS2} to \ref{Fig_WSe2}, we also may conclude that the renormalized mass has two main regimes in terms of its temperature dependence, namely, one for low temperatures, where $E_{\rm{opt}}$ is almost constant, and the second for higher temperatures, where  $E_{\rm{opt}}$ is well described by a linear decreasing behavior in terms of $T$. 
Interestingly, the bare mass $m$ works as a reference temperature separating these phases. Indeed, let us consider the ratio $\theta = m/k_B$. This temperature separates the curves into these two regions: one with a non-linear behavior ($T < \theta$) and the other with a linear behavior ($T > \theta$). The temperatures $\theta$ associated with each material are $\theta_{MoS_2} = 142.0$~K, $\theta_{MoSe_2} = 110.2$~K, $\theta_{WS_2} = 136.3$~K, and $\theta_{WSe_2} = 120.0$~K. Obviously, this is only a rough estimate for such a transition and it seems to be related to the fact that the phonon correction is actually more relevant in the regime $T\geq \theta$. 
On the other hand, the parameter $\bar{g}$ is related to the el-ph coupling constant, which is taken as a free parameter in order to compare with the experimental data, as it is shown in Table~\ref{Table_1}.

Note that the optical band gap agrees with the experimental data (within a certain range of temperatures), hence, the same conclusion holds for $2m^R(T,\Lambda)$, because they only differ by a small fraction (the binding energy $E_b$). 
It is important to highlight that our results provide an equation to fit the electronic band gap as well. Specifically, Eq.~\eqref{mr1} [such as Eq.~\eqref{mr2}] describes the behavior of the electronic band gap as a function of temperature. Although we do not explicitly explore the electronic band gap, its value can be obtained from Figs.~\ref{Fig_MoS2} to \ref{Fig_WSe2} using Eq.~\eqref{Eopt}. 

We can analyze the renormalized electronic band gap values in the low-temperature regime. In order to do so, we can extract the first data points in Figs.~\ref{Fig_MoS2}--\ref{Fig_WSe2}, which correspond to experimental measurements taken at $4.5$~K~\cite{Liu}. In this temperature, our renormalized mass is close to $M(0,\Lambda)$, therefore Eq.~\eqref{mr2} yields $m^R(T=4.5~{\rm{K}},\Lambda)~\approx~M(0,\Lambda)$ [see discussion below Eq.~\eqref{M0}]. Under this condition, Eq.~\eqref{Eopt} is rewritten as $E_{\rm{opt}}(T =4.5~{\rm{K}},\Lambda)=2M(0,\Lambda)-|E_{b}|$. The optical band gap for MoS$_2$ is $E_{\rm{opt}}(T =4.5~{\rm{K}}, \Lambda)=1.9$ eV (see Fig.~\ref{Fig_MoS2}), with an exciton binding energy of roughly $E_{b}=500.0$ meV \cite{Hill}. Consequently, the electronic band gap is approximately $2M(0,\Lambda)=2.4$ eV, which agrees with the experimental data of Ref.~\cite{Hill}. A similar analysis can be applied to other materials, whose renormalized mass values are in good agreement with experimental data \cite{Ugeda, Hanbicki, FLiu, Nguyen}. It is worth noting that these values depend on additional factors, such as screening effects and the density of charge carriers.


As we have mentioned before, our renormalized mass $m_R$ can be related to the band curvature around the $K$ point of these monolayers of TMDs. Next, for the conduction electron, we discuss the effective mass $M_{\rm {eff}}$, obtained from $\hbar^2/M_{\rm {eff}}=d^2E/dk^2$ at $k\rightarrow 0$, where $E(k)=\sqrt{\hbar^2 v^2_F k^2+c^4 m_R^2}$ is our energy dispersion. For the sake of simplicity, we have recovered physical units in order to compare with experimental findings. The $M_{\rm {eff}}$ represents the mass of a quasiparticle with a parabolic energy dispersion around the $K$ point (which we consider as our Dirac point $k\rightarrow 0$). Therefore, after a simple algebra, it follows that $M_{\rm {eff}}=m_Rc^2/v^2_F$. Let us call $m_0$ as the rest mass of the electron and define a ratio given by $r\equiv M_{\rm {eff}}/m_0$, which is an important parameter discussed in several works for monolayers of TMDs \cite{Nguyen, Aghajanian, Kadantsev, Sara, Jin, Goryca}.

\begin{table}[h]
            \centering
            \caption{The parameter $r$ is the ratio between the effective mass ($M_{\text{eff}}$) and the mass of the electron at rest ($m_0$). Below are shown the values of $r$ we calculated, as well as the theoretical and experimental values found in the literature.} \label{Table_2}
            \begin{tabular}{lccc}
            \hline
                Material & Our work &  Theor. & Exper. \\
                \hline
                MoS$_2$ & $ [0.38, 0.87]$ & $ [0.40, 0.54]$\textsuperscript{\tiny{\cite{Aghajanian, Kadantsev, Sara}}} & $ 0.7 \pm 0.1$ \textsuperscript{\tiny{\cite{Nguyen}}} \\
                MoSe$_2$ & $ [0.33, 0.75]$ & $ 0.43$ \textsuperscript{\tiny{\cite{Sara}}} & $ 0.5 \pm 0.1$ \textsuperscript{\tiny{\cite{Nguyen}}} \\
                WS$_2$ & $ [0.39, 0.89]$ & $ 0.33$ \textsuperscript{\tiny{\cite{Sara}}} & $ 0.5 \pm 0.1$ \textsuperscript{\tiny{\cite{Nguyen}}} \\
                WSe$_2$ & $ [0.33, 0.74]$ & $ 0.36$ \textsuperscript{\tiny{\cite{Sara}}} & $ 0.42 \pm 0.05$ \textsuperscript{\tiny{\cite{Nguyen}}}  \\
                \hline
            \end{tabular}
\end{table}

For monolayers of TMDs, we have an exact relation $E_{\rm{{opt}}}=-E_b+E_g$, where $E_b$ is the binding energy and $E_{\rm{{opt}}}$ is the optical band gap shown in the $y$-axis of Figs.~\ref{Fig_MoS2}--\ref{Fig_WSe2}. As an example, we roughly estimate the value of $r$ for MoS$_2$ with $E_b\approx 500$ meV. In the low-temperature limit ($T\approx 4$ K), we have $E_{\rm{{opt}}}\approx 1.96$ eV (see Fig.~\ref{Fig_MoS2}), hence, $E_g\approx 2.46$ eV and $c^2m_R \approx 1.23$ eV. Using a typical interval for the Fermi velocity of $v_F \in [c/600,c/400]$ (which represents $v_F\in [0.5,0.75]\times  10^6$ m/s, see Ref.~\cite{Xiao, Hao, Mitioglu}), we have $M_{\rm {eff}}=m_R c^2/v^2_F\in [0.20,0.44]$ MeV. Our estimated range for $v_F$ (between \(c/600\) and \(c/400\)) reflects experimental uncertainties and renormalization effects due to electron-phonon interactions and screening of charges. This explains the intervals for $M_{\text{eff}}$. Finally, the rest energy of the electron is $c^2m_0\approx 0.51$ MeV. Therefore, $r\equiv M_{\rm {eff}}/m_0 \in [0.38,0.87]$. Our estimative for $M_{\rm {eff}}$ is, therefore, quite sensitive to the precise value of $v_F/c$. Similar calculations are straightforward for other monolayers of TMDs. In Table~\ref{Table_2}, we compare our intervals for $r$ with recent theoretical and experimental works. From Table~\ref{Table_2}, we observe a partial agreement between previous results for $r$ and our estimated value. The main reason is that the precise values of key parameters such as $v_F$ and $E_b$ depend on the details of the sample, such as substrate and charge carrier density.

\section{Summary and Outlook}\label{Sec_SummOut}

The experimental realization of graphene in 2004 and its Dirac cones in the low-energy limit has built a critical point of intersection between the physics of high-energy theories, such as quantum electrodynamics, and two-dimensional materials. The realization of TMD monolayers increases these applications because the matter field can have a bare mass, generated by the buckled structure of a honeycomb lattice. In this case, the matter field is given by a massive Dirac-like equation in (2+1)D. The usefulness of these models has been clear due to both the calculation of the renormalized Fermi velocity and the electronic band gap, where the theoretical results are in agreement with the experimental findings. Obviously, in this approximation, the full dependence on the lattice $a$ is not considered and one must consider the continuum limit, i.e., $a\rightarrow 0$, which implies a low-energy regime. For considering the effects of $a\neq 0$ and the finite size of samples, it seems that the standard condensed matter models are more adequate. Nevertheless, whenever the continuum limit is reasonable, the quantum field theory approximation is very successful in including the quantum corrections and, therefore, calculating renormalized parameters that may be verified in experimental observations.

In this work, we have derived a theoretical expression of the renormalized optical band gap in monolayers of TMDs, within the low-energy limit that has been carefully discussed in Sec.~\ref{Sec_Intro} and Appendix \ref{App_Model}. This result has an excellent agreement with the experimental findings for temperatures ranging from $[4,500]$\,K. Furthermore, we also have considered that the el-ph interaction is small and performed the large-$N$ expansion. These strong conditions are needed to ensure that the dynamics of the phonon field remain much less relevant than the dynamics of the charge carriers (the Dirac-like field). This, however, seems quite physical because the velocity of the electrons is always much larger than the velocity of the phonons in these systems. We also prove that this condition implies that the vacuum polarization tensor corrections are suppressed. It is important to note that these experimental results have been well described by using phenomenological equations, which have been discussed since the 1950s. Our effective model, however, yields the correct expression for $m^R(T,\Lambda)$ from a straightforward one-loop quantum correction to the electron propagator, and it allows further generalization to other systems. Indeed, let us assume a system where electronic interactions can not be neglected. Hence, one should couple the Dirac matter to the gauge field of pseudo-quantum electrodynamics and obtain a model that describes both electron-electron and electron-phonon interaction. Furthermore, our model could also be generalized to a system where electrons are subjected to an external magnetic field considering the magnon-electron interaction or magnon-phonon interaction and include other many-body interactions such as impurities or disorders. We shall discuss these cases elsewhere.

\section*{Acknowledgement}
We would like to express our gratitude to Danilo T. Alves, E. C. Marino, and T. H. Hansson for their invaluable discussions, insightful suggestions, and constructive feedback. This work was partially supported by Conselho Nacional de Desenvolvimento Científico e Tecnológico (CNPq), Brasil, Processo 408735/2023-6 CNPq/MCTI. N. B. was partially supported by Coordenação de Aperfeiçoamento de Pessoal de Nível Superior Brasil (CAPES), finance code 001. 
L.F. acknowledges the financial support from Dirección de Investigación y Desarrollo de la Universidad de La Frontera through the Proyecto BPVRIP-14-2024.
\appendix
\numberwithin{equation}{section}
\section{Effective electron-phonon interaction }\label{App_Model}

In this appendix, we provide a more detailed explanation of the model introduced in Sec.~\ref{Sec_FeynmanRules}.

Having the electronic low-energy description in mind, let us discuss the lowest-order perturbative approach for both the phonon field and el-ph interaction when considering a lattice deformation. It turns out that the lattice vibrations change the atomic potential, generating the el-ph interaction \cite{Mousavi}. The relative displacement of two sublattices $A$ and $B$ is given by $\bold{u}(\bold r) = \bold{u}_A(\bold r) - \bold{u}_B(\bold r)$ \cite{Sasaki_1, Ando_1}, where $\bold{r}$ is any coordinate vector in the unit cell. Here, we would like to describe the optical modes at low energies. In order to do so, we consider the long-wavelength limit, $\bold{q} \to 0$, where $\bold{q}$ is the phonon momentum, corresponding to the $\Gamma$ point phonon mode. Within this picture, the relative displacement is given by \cite{Ando_1, Sasaki_3}
\begin{equation}
\bold{u}(\bold r) = \sum_{q,\lambda} \frac{e^{i \bold q.\bold r}}{\sqrt{2 N_c M \omega_0}} \left(b^{\dagger}_{-q,\lambda} + b_{q,\lambda} \right)   \bold{\hat{e}}_\lambda \, ,
\end{equation}
where $b^{\dagger}_{q,\lambda}$ and $b_{q,\lambda}$ are the creation and annihilation operators of the phonon field, respectively. The index $\lambda$ denotes either the longitudinal optical (LO) and in-plane transverse optical (iTO) modes, i.e., $\lambda= \{$LO, iTO$\}$, as it has been done in Ref.~\cite{Sasaki_1, Ando_1}, and $\omega_0$ is the phonon frequency within the long-wavelength approximation. Furthermore, $\bold{\hat{e}}_\lambda$ is the polarization of the phonon field, $N_c$ is the number of unit cells, and $M$ is the ion mass.
 
In this work, we describe the interaction between the long-wavelength phonon modes with electrons (and holes) close to the Dirac point. 
These phonons are predictable around this point \cite{Giner} and measurable using inelastic spectroscopy \cite{Tornatzky}, attributed to their dispersion along the $\Gamma$-$K$ direction of the Brillouin zone \cite{Campos, Giner}. 
In this case, a lattice deformation induces a local modification of the nearest-neighbor hopping integral, as it has been shown in Refs.~\cite{Ando_1, Sasaki_1}. We are, therefore, considering independent electrons and neglect the Coulomb interaction between them. Hence, the resulting el-ph interaction generated near the $K$ and $K'$ points is given by \cite{Sasaki_1, Neto2009}
\begin{equation}
\mathcal{H}_{int} = \sum_{\mathtt{a}=1}^N \int d^2	 \bold{r}  \, \psi_{\mathtt{a}}^{\dagger}(\bold r)  \pmb{\sigma} \,.\, \pmb{\mathcal{A}} (\bold r)  \psi_{\mathtt{a}}(\bold r), \label{Hint}
\end{equation}
where $\pmb{\mathcal{A}} =(\mathcal{A}_x, \mathcal{A}_y)$ is called as \textit{deformation-induced gauge field} \cite{Sasaki_1, Sasaki_2, Ando_1, Ando_2, Neto2009}, which is connected to the displacement through the relation $(\mathcal{A}_x, \mathcal{A}_y) = (\alpha/a \sqrt{3}) (u_y, - u_x)$, where $\alpha$ is the el-ph coupling strength with units of energy, which we shall take as a free parameter and $a$ is the lattice parameter. 

In this framework, we can describe the phonon model in a crystal using a set of coupled harmonic oscillators \cite{Kittel, Ashcroft} written in terms of $\bold{u}(\bold r)$ or $\pmb{\mathcal{A}}(\bold r)$, as primary variables, and their derivatives. However, we shall consider a further approximation, namely, the Born-Oppenheimer approximation \cite{Zhou, Feliciano}, also known as the adiabatic approximation, because $\omega_0/t \ll 1 $, where $t$ is the hopping integral \cite{Zhou}. This is a reasonable approximation whenever the velocity of the charge carriers is much greater than the phonon velocity, such that the phonon dynamics can be neglected and only a term proportional to $u_x^2$ and $u_y^2$ are considered. Moreover, we have only one vibrational mode $\omega_0$. 
Furthermore, it is known that in these systems, the quasiparticles near the Dirac point are described by the Dirac equation \cite{MarinoBook, Neto2009}. 
This description considers only the first-neighbor hopping, where charge conjugation symmetry is preserved, and it is expected to work for energies close to the high-symmetry points $K$ and $K'$ in the first Brillouin zone. 
In this context, the effective action for our model in Euclidean space is expressed as \cite{Andreas2016, Andreas2020}:
\begin{equation}
\begin{split}
	S_{\rm{eff}} = \sum^N_{\mathtt{a}=1} \int d \tau \int d^2 \bold{r} &\left[ \bar{\psi}_\mathtt{a} \left(i \gamma^{\mu} \partial_{\mu} - m \right)\psi_\mathtt{a} \right. \\
	  &-\left. \frac{1}{2 \,g} \mathcal{A}^i \mathcal{A}_i + \bar{\psi}_\mathtt{a} \gamma^i  \mathcal{A}_i \psi_\mathtt{a} \right]\, , 
\end{split} \label{Seff}
\end{equation}
where $g = (\alpha/\sqrt{3 \rho} \, a \, \omega_0)^2$ is the coupling constant, which works as our free parameter, and $\rho$ is the ion mass density. $\gamma^\mu = (\gamma_0,\gamma_1,\gamma_2) =(i\sigma_z,i\sigma_y,-i\sigma_x)$ are our Dirac matrices in (2+1)D, where $\mu=\{ 0,1,2\}$ and the index of the phonon field is $i=\{1,2\}$. Finally, $\bar{\psi} = i\psi^{\dagger} \gamma^0$ is the adjoint spinor.
 
From a theoretical perspective, the vector field $\mathcal{A}_i$ is an auxiliary field that emerges after the Hubbard-Stratonovich transformation \cite{Stratonovich, Hubbard} is applied to the Thirring model \cite{MGomes}. 
Interestingly, this interaction has been largely discussed in the realm of high-energy physics in quantum field theory \cite{Hong, Hands, Itoh, Gies, Gehring, MGomes}. In our case, however, the Thirring-like interaction has an O(2) symmetry and the term $(\bar\psi\gamma_0\psi)^2$ vanishes \cite{Andreas2016}. The main consequence, therefore, is that it is possible to include a real two-component vector field $ \mathcal{A}_1$ and $ \mathcal{A}_2$, coined as the phonon field, in order to describe the el-ph interaction \cite{Andreas2020}.

The discussions so far cover all of the model-based approximations we have considered, motivated by physical arguments. 
In Sec.~\ref{Sec_FeynmanRules} we apply the so-called large-$N$ expansion and rewrite the trilinear interaction by substituting $\bar{\psi}_\mathtt{a} \gamma^i \mathcal{A}_i \psi_\mathtt{a} \to \bar{\psi}_\mathtt{a} \gamma^i \mathcal{A}_i \psi_\mathtt{a}/\sqrt{N}$ [see Eq.~\eqref{L_Model}].

\section{Some important limits}\label{App_Limits}

In this appendix, we calculate the zero ($T\rightarrow 0)$ and high ($T\rightarrow\infty$) temperature limit of Eq.~(\ref{mr1}). In order to do so, let us define two positive real constants, namely, $a=(\Lambda^2+m^2)^{1/2}$ and $b=m$.

\subsection{The low-temperature regime}\label{Sub_App_T0}
Firstly, let us take $T \to 0$ in Eq.~(\ref{mr1}). Hence, the second right-hand side term of this equation may be simplified as
\begin{equation}
\begin{split}
f(T \to 0) &= \lim_{T \to 0} T \ln \left[ \frac{\cosh(a/T)}{\cosh(b/T)} \right] \\
&= \lim_{T \to 0} T \ln \left[ \frac{e^{a/T} (e^{-2a/T}+1) }{e^{b/T}(e^{-2b/T}+1)} \right]\\
&=  \lim_{T \to 0} T \left\lbrace \frac{(a-b)}{T} + \ln \left[ \frac{ e^{-2a/T}+1 }{e^{-2b/T}+1} \right] \right\rbrace \\
&= a-b.
\end{split}
\end{equation}
This result can be used to promptly derive Eq.~(\ref{M0}).

\subsection{The high-temperature regime}\label{Sub_App_Tinfinity}
Next, let us consider $T \to \infty$ in Eq.~(\ref{mr1}). Hence, the second right-hand side term of this equation is now written as
\begin{equation}
\begin{split}
g(T\to \infty) &= \lim_{T \to \infty} T \ln \left[ \frac{\cosh(a/T)}{\cosh(b/T)} \right] \\
&= \lim_{T \to \infty} \frac{\ln \left[ \frac{\cosh(a/T)}{\cosh(b/T)} \right]}{1/T}, \\
\end{split}
\end{equation}
which is an indeterminate limit. However, using the L'Hospital's rule, we find
\begin{equation}
\begin{split}
g(T\to \infty) =& \lim_{T \to \infty} \left[ \frac{b \,\tanh(b/T)}{T^2 } -  \frac{a \,\tanh(a/T)}{T^2 }  \right] \\
&\times  \left( \frac{1}{-1/T^2} \right)  \\
=& \lim_{T \to \infty}  \left[ a \, \tanh(a/T)  - b \, \tanh(b/T) \right] \\
=& \, 0 \,.
\end{split}
\end{equation}
Therefore, from Eq.~(\ref{mr1}), we conclude that $ m^R(T,\Lambda) \rightarrow m $ whenever $T\gg\Lambda$. This result has a simple physical interpretation, it only means that the quantum correction vanishes at a very high temperature, hence, the mass remains at its bare value. For a real two-dimensional material, this threshold temperature would be close to the decomposition temperature which is in order of $10^3$\,K. 

\section{Electron self-energy for $T=0$}\label{App_SEwithoutTemp}
In this appendix, we solve Eq.~\eqref{SE_T0} for the case $T=0$. After solving for the Dirac matrices product and using $d^2 k = \mathrm{k}\, d\theta \, d\rm k$ with $|\bold k | = \rm k$, $|\bold p | = \rm p$, $\mathrm k  \to \mathrm{(k+p)}/v_F$, and $\mathrm p  \to\mathrm p/v_F$, we have
\begin{equation}
\Sigma (\Lambda) = - \frac{\bar{g} \, m}{2} \int^\Lambda_0 d\, \mathrm k \frac{\rm k}{(\mathrm k^2 + m^2)^{1/2}} \,,
\end{equation}
\n where we have defined $\bar{g}=g/N\pi v^2_F$.
This integral is solved in Eq.~\eqref{Inte1}, yielding the electron self-energy, namely,
\begin{equation}\label{SE_Solv_T0}
\Sigma (\Lambda) = - \frac{\bar{g}}{2} m \left[ (\Lambda^2 + m^2)^{1/2} - m \right] \, .
\end{equation}
Therefore, after using the Schwinger-Dyson equation, we find the renormalized mass given by
\begin{equation}\label{mr0}
m^R(\Lambda) = m + \frac{\bar{g}}{2} m \left[ (\Lambda^2 + m^2)^{1/2} - m \right] \,.
\end{equation}
Note that Eq.~\eqref{mr0} is the same as Eq.~\eqref{M0}, as expected. 
              

\end{document}